\documentclass[twocolumn,secnumarabic,amssymb,aps,prl,superscriptaddress]{revtex4}

\usepackage{graphicx}
\usepackage{enumerate}
\usepackage{color}

\begin{document}
\title{ Error- and Loss-Tolerances of Surface Codes with General Lattice Structures}

\author{Keisuke Fujii}
\affiliation{Graduate School of Engineering Science, Osaka University,
Toyonaka, Osaka 560-8531, Japan}
\author{Yuuki Tokunaga}
\affiliation{ NTT Information Sharing Platform Laboratories, NTT Corporation, 3-9-11 Midori-cho, Musashino, Tokyo 180-8585, Japan}
\affiliation{Japan Science and Technology Agency, CREST, 5 Sanban-cho, Chiyoda-ku, Tokyo 102-0075, Japan}

\begin{abstract}
We propose a family of surface codes with general lattice structures,
where the error-tolerances against bit and phase errors 
can be controlled asymmetrically by changing the underlying lattice geometries. 
The surface codes on various lattices 
are found to be efficient in the sense that
their threshold values universally
approach the quantum Gilbert-Varshamov bound.
We find that the error-tolerance of surface codes
depends on the connectivity of underlying lattices;
the error chains on a lattice of lower connectivity are
easier to correct.
On the other hand, the loss-tolerance of surface codes 
exhibits an opposite behavior;
the logical information on a lattice of higher connectivity
has more robustness against qubit loss.
As a result, 
we come upon a fundamental trade-off between 
error- and loss-tolerances in the family of the surface codes with different lattice geometries.

\end{abstract}

\maketitle

{\it Introduction.---}
Recently, topological order has attracted much interest 
in both condensed matter physics \cite{Wen,TopologicalQC} 
and quantum information science \cite{Kitaev}.
The ground state degeneracy of topologically ordered phase
cannot be distinguished by local operations and hence robust against local perturbations.
By encoding quantum information
into such topologically degenerate subspaces, 
so-called topological quantum error correction (QEC) codes \cite{Kitaev,Bombin06},
logical information can be protected from decoherence 
by repeated quantum error correction.
There have been two types of the topological QEC codes, so-called surface codes \cite{Kitaev} and color codes \cite{Bombin06},
both of which are the CSS (Calderbank-Shor-Steane) codes \cite{CSS}.
In both cases, the threshold values under perfect syndrome 
measurements have been calculated to be $\sim 11\%$ \cite{Denis,Katzgraber}, 
which is close to the quantum Gilbert-Varshamov bound \cite{CSS}
in the limit of zero asymptotic rate with symmetric $X$ and $Z$ errors.
In the case of the color codes,
their performances have been compared among
different lattice geometries,
and their thresholds result in similar values $\sim 11\%$ \cite{Katzgraber,Ohzeki}.
This result is reasonable by considering the fact that the color codes
are self-dual CSS codes,
that is, they are symmetric under a Hadamard transformation.
The surface codes, on the other hand, are not self-dual CSS codes,
and hence it is possible to break the symmetry between properties 
of $X$ and $Z$ error-corrections.
The surface code, however, has been intensively investigated so far
only on the square lattice, which is a self-dual lattice, and therefore
its error-correction properties are symmetric.

In this letter,
we investigate the surface codes with general lattice geometries.
Their constructions and error-correction procedures are 
basically the same as those of the original surface code.
Since the stabilizer operators are not always symmetric under the duality transformation
of the lattice (i.e. exchange of the vertexes and faces with each other),
the error-tolerances of the surface codes with general lattice geometries
are not always symmetric between $X$ and $Z$ errors.
Interestingly, we find that
such asymmetry in the error-tolerance is related to the connectivity 
of the lattice which defines the surface;
{\it error chains on a lattice of lower connectivity
can be corrected easily}.
Intuitively, this can be understood that finding appropriate pairs of incorrect error syndromes,
which are the boundaries of the error chains, 
on the lattice of lower connectivity
is easier, since the incorrect error syndromes are more isolated and less percolative.
Furthermore,
we find that the threshold values for independent $X$ and $Z$ errors
exhibit a universal behavior;
they, independent of the lattice geometries, 
approach the quantum Gilbert-Varshamov bound \cite{CSS}
in the limit of zero asymptotic rate with asymmetric $X$ and $Z$ errors.
In this sense, the family of the surface codes can be said to be efficient.
We also provide a recursive way to construct 
highly asymmetric surface codes on fractal-like lattices.
In many experimental situations,
dephasing is a dominant source of errors \cite{Aliferis},
and therefore
the present family of asymmetric surface codes
will help us to correct such biased noise efficiently.

The loss-tolerant scheme 
is based on a bond percolation phenomenon,
where a reliable logical operator can be reconstructed 
even on the lossy surface as long as the survival probability of the qubits
is higher than the bond percolation threshold \cite{Barrett}.
Thus {\it the logical information on a lattice of higher connectivity
is robust against qubit loss}.
As a result, we come upon a fundamental trade-off between
error- and loss-tolerances of the surface codes depending on the
connectivity of the underlying lattices;
the logical information on a lattice of higher connectivity
is robust against qubit loss, but
the error chains on such a lattice are difficult to correct (and vice versa).
It is interesting to note that such a trade-off between 
error- and loss-tolerances has been also discussed 
in a far different situation \cite{Rohde}.

\begin{figure}
\centering
\includegraphics[width=80mm]{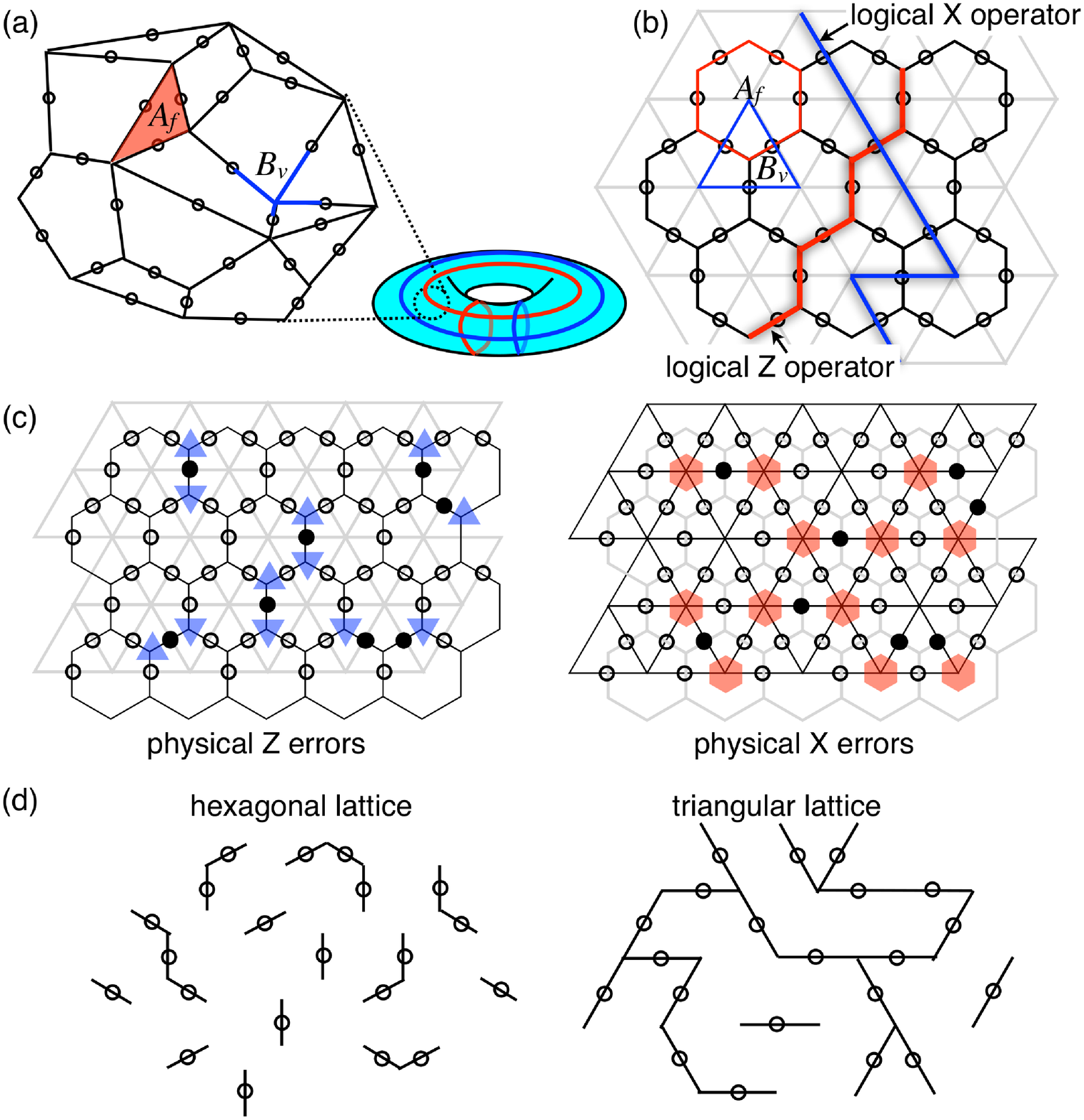}
\caption{ (Color online) (a) A surface code with a general lattice structure.
(b) The surface code on the hexagonal lattice and the logical operators. (c) 
Incorrect error syndromes against $Z$ (left) and $X$ (right)
errors, which are associated with the chains on
the primal and dual lattices respectively.
Here, the $Z$ (left) and $X$ (right) errors are located 
at the same qubits depicted by the solid circles. 
(d) The remaining edges (qubits) on the primal (left) and dual (right) lattices
after qubit loss. The logical $Z$ and $X$ operators 
are associated with the non-contractible loop on the primal and dual lattices,
respectively.
}
\label{Fig1}
\end{figure}

{\it Surface codes on general lattices.---}
Let us consider a lattice $\mathcal{L}(E,V,F)$,
where $E$, $V$, and $F$ are the sets of edges, vertices, and faces of the lattice,
respectively [see Fig. \ref{Fig1} (a)].
A qubit is associated with each edge.
The stabilizer operators of the surface code on the lattice $\mathcal{L}$
are defined for each face $f \in F$ and vertex $v \in V$,
respectively, as 
\begin{eqnarray*}
A_f^{\mathcal{L}}=\bigotimes _{ i \in E_f } Z_{i} ,
\;\;\;\;
B_v^{\mathcal{L}}=\bigotimes _{j \in E_v} X_j.
\end{eqnarray*}
Here, $Z_i$ and $X_j$ denote
Pauli operators on the $i$th and $j$th qubits ($i,j \in E$) respectively,
and $E_{f}$ and $E_v$ ($E_{f,v} \subset E$) indicate the sets of edges 
which are surrounding the face $f$ and are adjacent to the vertex $v$, respectively.
From these definitions we define the lattice $\mathcal{L}$ as primal.
Its dual lattice, which is denoted by $\bar {\mathcal{L}}(\bar E, \bar V, \bar F)$,
can be defined by exchanging the vertices and faces of $\mathcal{L}$ with each other. 
Similarly to the original surface code on the square lattice,
$Z$ and $X$ errors are detected as incorrect error syndromes
at boundaries of $Z$ and $X$ error chains on the edges of primal and dual lattices,
respectively.
The logical $Z$ and $X$ operators are defined as tensor products of Pauli $Z$ and $X$ operators
on the noncontractible loops on the edges 
of the primal and its dual lattices [see Fig. \ref{Fig1} (b)],
respectively.
In the following, the logical operators play the following two roles:
(i) If the error correction is failed, a logical operator
acts on the code space, which we call  {\it a logical error}.
(ii) The logical operators represent encoded quantum information,
which we call {\it logical information}.
Since $H^{\otimes |E|} A_f^{\mathcal{L}} H^{\otimes |E|} 
= B_{\bar v}^{\bar {\mathcal{L}}}$,
the surface code defined on the primal lattice $\mathcal{L}$ 
is equivalent to that defined on the dual lattice $\bar {\mathcal{L}}$  
up to the Hadamard transformation $H^{\otimes |E|}$.
In the case of the original surface code,
which is defined on the square lattice, 
the $Z$ and $X$ error-tolerances
are symmetric, since $\mathcal{L} = \bar{\mathcal{L}}$ up to translation.

Let us consider, for instance, a surface code 
on the hexagonal lattice as shown in Fig. \ref{Fig1} (b).
In such a case, $\mathcal{L} \neq \bar{\mathcal{L}}$, and hence
the $Z$ and $X$ error-tolerances
are no longer symmetric.
Suppose that $Z$ or $X$ errors are, for example, 
located at those qubits depicted with solid circles in Fig. \ref{Fig1} (c).
The $Z$ and $X$ errors are detected at the boundaries (vertices) of the error chains 
on the edges of the hexagonal (primal) and triangular (dual) lattices, respectively,
which are denoted by blue (left) triangles and red (right) hexagons in Fig. \ref{Fig1} (c).
In the error-correction procedure, we have to infer the locations of the errors
from the information of incorrect error syndromes.
As can be seen clearly from Fig. \ref{Fig1} (c),
$X$ errors (right) are less distinguishable than $Z$ errors (left),
since incorrect error syndromes are more dense and percolative.
More precisely,
the error syndromes of $Z$ and $X$ errors are associated with
the vertices on the primal and dual lattices, respectively,
and therefore the distinguishability of the $Z$ and $X$ errors depends on
the connectivity of the primal and dual lattices, respectively.
The bond-percolation threshold is one of the most well-known measures of 
the connectivity of lattices.
The bond-percolation thresholds of the 
hexagonal (primal) and triangular (dual) lattices 
are given by $0.653$ and $0.347$, respectively.
Since the hexagonal (primal) lattice has lower connectivity, 
the surface code on the hexagonal lattice 
is expected to be more robust against $Z$ errors.

\begin{table*}
\begin{center}
\begin{tabular}{c|c|c|c|c}
\hline \hline
lattice  &   $\;\;\; p_Z^{\rm th} \;\;\;$ & bond-percolation (primal) & $\;\;\; p_X^{\rm th} \;\;\; $ & bond-percolation (dual) 
\\
\hline
square &  0.103 & 1/2 &  0.103 & 1/2
\\
Kagome  & 0.116 & 0.524 & 0.095& 0.476 
\\
hexagonal  & 0.159 & 0.653 & 0.065 & 0.347
\\
tri-hexa & 0.205 & 0.740 & 0.041  & 0.260  
\\
\hline \hline 
\end{tabular}
\end{center}
\caption{Summary of the threshold values of the surface codes on various
 lattices. 
The bond-percolation thresholds of the primal and dual lattices are also shown.}
\label{threshold}
\end{table*}%

{\it Threshold values.---}
In order to confirm the above observation, we perform numerical simulations
and estimate the threshold values of the surface codes 
with various lattice geometries such as the
Kagome, hexagonal, and $(3,12^2)$ 
({\it triangle-hexagonal}), say {\it tri-hexa}, lattices.
For independent bit and phase errors 
with probability $p_X$ and $p_Z$, respectively,
topological error correction is simulated by using
the minimum-weight-perfect-matching algorithm.
Then, by using the finite-size scaling ansatz for
the logical error probability, $p^{l}_{Z,X} = A + B(p_{Z,X} - p^{\rm th}_{Z,X})L^{1/\nu}$ 
with fitting parameters $(A,B,\nu,p_{Z,X}^{\rm th})$ similarly to Ref.  \cite{Wang03},
the threshold values  $p_Z^{\rm th}$ and $p_X^{\rm th}$
against physical $Z$ and $X$ errors are calculated, respectively
 \cite{comment,comment_num}.
The resultant threshold values are summarized in Table \ref{threshold}.
As we expected, when the lattice is not self-dual,
the threshold values of the surface code
exhibit asymmetry between $Z$ and $X$ errors, and 
their behavior under various lattices can be understood from
the connectivity of the underlying lattices;
if a primal lattice has lower connectivity
(i.e. higher bond-percolation threshold),
the corresponding surface code is robust against $Z$ errors
(i.e. higher threshold value $p_Z^{\rm th}$). 
On the other hand, in such a case, the dual lattice exhibits higher connectivity
due to the Kesten's duality theorem 
(i.e. the sum of bond-percolation thresholds of a primal and its dual lattices is equal to one) 
\cite{Kesten}.
Thus the threshold value of the $X$ errors becomes lower \cite{comment5}.
Although there is no rigorous correspondence between
the bond-percolation model and the topological error correction,
it can be used as a good measure of the performance 
of the surface code.

\begin{figure}
\centering
\includegraphics[width=60mm]{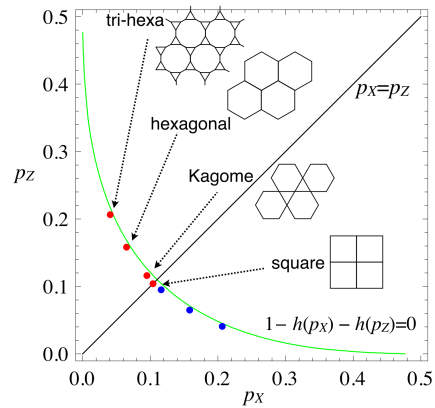}
\caption{(Color online) The threshold values $(p_X^{\rm th},p_Z^{\rm th})$ for 
the square, Kagome, hexagonal, and tri-hexa lattices are plotted (red dots) with 
the quantum Gilbert-Varshamov bound (green line). The blue dots are
the threshold values for the surface codes defined
on the duals of the Kagome, hexagonal, and tri-hexa lattices
(from left to right).}
\label{GVbound}
\end{figure}

In Fig. \ref{GVbound},
these threshold values $(p_X^{\rm th},p_Z^{\rm th})$
are plotted with the quantum Gilbert-Varshamov bound $R= 1 - h(p_X) - h(p_Z)$
in the limit of zero asymptotic rate $R\rightarrow 0$,
where $h(p)$ is the binary Shannon entropy \cite{CSS}.
Interestingly, for all lattices considered here, 
the threshold values $(p_X^{\rm th},p_Z^{\rm th})$ 
exhibit a universal behavior:
they, independent of the lattice geometries, approach the bound.
In this sense, the present family of the surface codes can be said to be efficient.
Furthermore, the scaling exponents are obtained to be $\nu \approx 1.5$
for all lattices, which indicates these surface codes
belong to the same universality class as that on the square lattice \cite{Wang03}.
From the above observation and these numerical evidences, 
it is expected in general that 
there is a trade-off between $X$ and $Z$ error-tolerances 
of the surface codes with other lattice geometries,
and the trade-off is well characterized by the quantum Gilbert-Varshamov bound.
Since the minimum-weight-perfect-matching algorithm is a sub-optimal decoding method,
the threshold values will be improved by using more sophisticated decoding algorithms
\cite{Barrett,Poulin}.

There is an exact correspondence between 
the threshold values $p^{\rm th}$ of the surface code and 
the critical point of the random-bond Ising model on the Nishimori line $e^{ -2 \beta _c }=p^{\rm th}/(1-p^{\rm th})$
where $\beta _c$ being the critical inverse temperature \cite{Kitaev}.
Thus, if the threshold value $p_{Z}^{\rm th}$ for the $Z$ errors is higher,
the corresponding random-bond Ising model on the primal lattice exhibits
a higher critical temperature.
On the other hand, in such a case, the threshold value $p_{X}^{\rm th}$
for the $X$ errors is lower,
and hence the critical temperature of the random-bond Ising model
on the dual lattice becomes lower.
According to the above result, 
the trade-off between the critical temperatures 
of the primal and dual lattices are also
characterized by the Gilbert-Varshamov bound 
through the Nishimori temperature
$e^{ -2 \beta _c }=p^{\rm th}/(1-p^{\rm th})$.

{\it Loss-tolerance.---}
Next we consider the performance of the surface codes
against qubit loss.
Recently, a novel loss-tolerant scheme has been proposed for the surface code on the square lattice \cite{Barrett}.
There, even if some qubits are lost, the stabilizer operators 
and the logical operators
can be reconstructed as long as the loss rate is less than a threshold.
The threshold value of qubit loss without errors is 
given exactly by the bond-percolation threshold of the lattice.
This is because one can reconstruct a logical operator
as long as the remaining qubits (bond) percolate through the lossy surface [see Fig. \ref{Fig1} (d)].
We can also apply the loss-tolerant scheme 
for the surface codes on the general lattices in a similar manner as in Ref. \cite{Barrett}.
Then, the loss-tolerances of the logical $X$ and $Z$ information 
also become asymmetric depending on underling lattice geometries.
According to the Kesten's duality theorem,
there is a trade-off between the bond-percolation thresholds of the primal and dual lattice.
Thus, asymmetry in the tolerable loss rates $p_{X}^{\rm loss}$ and $p_{Z}^{\rm loss}$ 
of the logical $X$ and $Z$ information, respectively,
is subject to a trade-off, $p^{\rm loss}_X +p^{\rm loss}_Z=1$. 

\begin{figure}
\centering
\includegraphics[width=80mm]{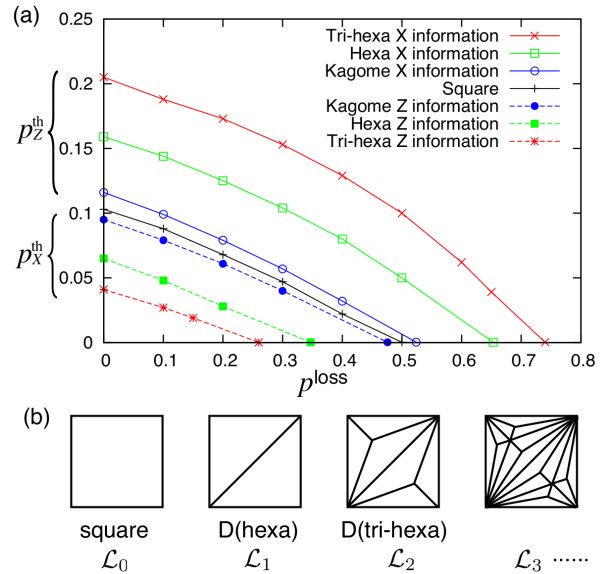}
\caption{(Color online) (a) The threshold curves $(p^{\rm loss},p^{\rm th})$ for
logical $X$ (solid line) and $Z$ (dotted line) information \cite{comment_loss}.
Top to bottom: tri-hexa ($X$), hexagonal ($X$), Kagome ($X$), square ($X$ and $Z$), Kagome ($Z$), hexagonal ($Z$), and tri-hexa ($Z$). (b) A recursive way to construct a fractal-like lattice $\mathcal{L}_{n}$ of higher connectivity. Its dual lattice $\bar \mathcal{L}_{n}$ has lower connectivity.}
\label{Loss}
\end{figure}

We also perform numerical simulations for the loss-tolerant scheme
on the Kagome, hexagonal, and tri-hexa lattices.
The resultant threshold curves $(p^{\rm loss}, p^{\rm th})$ of the error and loss rates
are plotted in Fig. \ref{Loss} (a) for each lattice for logical $X$ and $Z$ information.
The logical $X$ information is robust against both $Z$ errors and qubit loss
as shown in Fig. \ref{Loss} (a),
while the logical $Z$ information is fragile against both $X$ errors and qubit loss
\cite{comment5}.
These results can be understood from
the two trade-offs, which we have studied above,
the trade-off between the $Z$ and $X$ error-tolerances
[points on the vertical axis in Fig. \ref{Loss} (a)],
and the trade-off between the loss-tolerances of logical $Z$ and $X$ information
[points on the horizontal axis in Fig. \ref{Loss} (a)].
If we choose the primal lattice $\mathcal{L}$ of low connectivity
so that the surface code is robust against $Z$ errors
[see Fig. \ref{Fig1} (c) left],
then the bonds of such a primal lattice are less percolative,
and hence the logical $Z$ information 
becomes fragile against loss [see \ref{Fig1} (d) left].
This is also the case for fragility against $X$ errors and robustness of the logical $X$ information [see Fig. \ref{Fig1} (c) right and (d) right].
That is, there is a fundamental relationship between
robustness against errors and 
fragility of the logical information against qubit loss 
depending on the lattice geometries of the surface codes. 

{\it Systematic construction of asymmetric lattices.---}
In the above case, qubit loss occurs randomly according to 
the loss rate $p^{\rm loss}$.
It is also suggestive to consider a systematic qubit loss.
If we introduce a systematic qubit loss for the triangular (dual of the hexagonal) lattice,
whose threshold value is $15.9\%$ as shown in Table \ref{threshold},
then we can reconstruct the surface as the square lattice,
whose threshold value is $10.3\%$ 
[see Fig. \ref{Loss} (b) $\mathcal{L}_{0}$ and $\mathcal{L}_{1}$].
By doing the inverse operation of the systematic qubit loss,
namely, a systematic injection of qubits,
we can construct a new lattice $\mathcal{L}_{n}$ recursively, 
which has high connectivity (its dual lattice $\bar \mathcal{L}_{n}$ has low connectivity), 
as shown in Fig. \ref{Loss} (b).
Therefore,
the threshold value for the $Z$ ($X$) errors 
on $\bar \mathcal{L}_{n}$ ($ \mathcal{L}_{n}$), 
which has lower (higher) connectivity,
is expected to be higher (lower) than that of $\bar \mathcal{L}_{n-1}$
($ \mathcal{L}_{n-1}$).
On the other hand, the logical $Z$ ($X$) information is more fragile (robust) against qubit loss.

{\it Conclusion and discussion.---}
We have investigated the surface codes with general lattice geometries.
These surface codes are efficient, since
their threshold values approach the quantum Gilbert-Varshamov bound universally.
We have found that there are interesting relationships between 
error or loss-tolerances of the surface codes
and the geometrical properties of the underlying lattices,
where the connectivity of the lattice 
has an opposite effect in error and loss-tolerances.
This results in a fundamental trade-off between
robustness against errors and 
fragility of the logical information against qubit loss (and vice versa)
among the surface codes on the different lattices.
We also provide a recursive way to construct a highly asymmetric surface code
on a fractal-like lattice.
These results suggest that we can systematically find an efficient surface code
by selecting the underlying lattice geometry 
depending on the degree of biased noise.

The surface codes with general lattice geometries 
also support topologically protected measurement-based quantum computation 
by braiding the defects \cite{Raussendorf2D,Raussendorf3D}. 
The error- and loss-tolerances in fault-tolerant topological quantum
computation on the general lattices are an intriguing future problem.

It is an interesting correspondence
that a trade-off between error- and loss-tolerances
has been found also in a far different situation \cite{Rohde},
where lost information is recovered by using 
counter-factual indirect measurements on a specific shape of the cluster states.
In such a case, if we introduce more redundancy for the loss-tolerance, more errors are accumulated, which results in a similar trade-off.
This correspondence suggests that there might be more fundamental 
principle between error- and loss-tolerances;
{\it for two mutually noncommuting observables, if one observable is robust against qubit loss,
the other one must be fragile against environmental perturbations}
\cite{comment3}.

\begin{acknowledgments}
KF is supported by MEXT Grant-in-Aid for Scientific Research on Innovative Areas 20104003.
\end{acknowledgments}

{\it Note added.---}
During preparation of this manuscript \cite{QIP},
a related work has been appeared \cite{Rothlisberger},
which also supports generality of the present results.


\begin{thebibliography}{99}
\bibitem{comment}
Here, the logical error probabilities $p^{l}_{Z,X}$
mean that after the recovery operations logical $Z$ and $X$
operators act on the code space with probabilities $p^{l}_{Z,X}$, respectively.
If a logical $Z$ ($X$) error occurs,
the logical $X$ ($Z$) information 
is flipped,
since the logical operators are anticommute.
If $p< p_{Z}^{\rm th}$ ($p< p_{X}^{\rm th}$),
the probability $p^{l}_Z$ ($p^{l}_X$) of such a logical $Z$ ($X$) error
is suppressed rapidly by increasing the
lattice size $L$.
\bibitem{comment_num}
The numerical simulations are performed on the $L \times L$ lattice
with $L=16,24,32$ for the square lattice.
For other structures,
the lattices of the similar numbers of qubits are used.
\bibitem{comment5}
It is, of course, possible to construct
surface codes, which are robust against
$X$ errors, by defining their stabilizer operators
on the duals of the Kagome, hexagonal, and tri-hexa lattices
as shown in Fig. \ref{GVbound}.
This is also the case for the loss-tolerance,
where the logical $Z$ information becomes robust against qubit loss
in the surface codes defined on these duals.
\bibitem{comment_loss}
When $p^{\rm loss}$ is near the bond percolation threshold, 
the numerically obtained $p^{\rm th}$ 
lies below the drawn curve (not shown in Fig. \ref{Loss}), 
because the finite size effects become dominant, 
and therefore the universal scaling law breaks down
similarly to the case of the square lattice \cite{Barrett}.
\end{thebibliography}
\end{document}